\documentclass[pra,nobalancelastpage,twocolumn,superscriptaddress,nofootinbib]{revtex4}

\usepackage{graphicx}
\usepackage{amssymb}
\usepackage{amsmath}
\usepackage[english]{babel}
\usepackage{color}
\usepackage[version=4]{mhchem}

\newcommand{\PMA}{Division of Physics, Mathematics and Astronomy, California Institute of Technology, Pasadena, CA 91125, USA}

\newcommand{\AP}{Thomas J. Watson, Sr., Laboratory of Applied Physics, California Institute of Technology, Pasadena, CA 91125, USA}

\newcommand{\WI}{Department of Physics, University of Wisconsin-Madison, 1150 University Avenue, Madison, WI 53706, USA}

\newcommand{\ket}[1]{ | #1 \rangle}
\newcommand{\bra}[1]{\langle #1 |}

\begin{document}
\title{Microwave-to-optical conversion via four-wave-mixing in a cold ytterbium ensemble}
\author{Jacob P. Covey}\email{covey@caltech.edu}
\affiliation{\PMA}
%\affiliation{\IQIM}
\author{Alp Sipahigil}
%\affiliation{\IQIM}
\affiliation{\AP}
%\author{Neil Sinclair}
%\affiliation{\PMA}
%\affiliation{\IQIM}
%\affiliation{\AQT}
%\author{Manuel Endres}
%\affiliation{\PMA}
%\affiliation{\IQIM}
%\author{Oskar Painter}
%\affiliation{\IQIM}
%\affiliation{\AP}
\author{Mark Saffman}
\affiliation{\WI}

\begin{abstract}
Interfacing superconducting qubits with optical photons requires noise-free microwave-to-optical transducers, a technology currently not realized at the single-photon level. We propose to use four-wave-mixing in an ensemble of cold ytterbium (Yb) atoms prepared in the metastable `clock' state. The parametric process uses two high-lying Rydberg states for bidirectional conversion between a 10 GHz microwave photon and an optical photon in the telecommunication E-band. To avoid noise photons due to spontaneous emission, we consider continuous operation far detuned from the intermediate states. We use an input-output formalism to predict conversion efficiencies of $\approx50\%$ with bandwidths of $\approx100$ kHz.
\end{abstract}

\maketitle

\section{Introduction}
Superconducting circuits are a promising candidate for realizing scalable quantum computation~\cite{Devoret2013,Barends2014}. However, their operation at microwave frequencies precludes their integration in a long-distance quantum network~\cite{Kimble2008}, which requires optical photons in the telecommunication wavelength window ($\sim1.25-1.65$ $\mu$m). Therefore, a coherent, low-noise, high-bandwidth quantum transducer between the microwave and optical domains that is both reversible and can operate at the typical single-photon bandwidths of superconducting circuits is desirable.

There are multiple prominent approaches to coherent microwave-to-optical conversion (MOC) that use mechanical resonators~\cite{SafaviNaeini2011,Higginbotham2018,Bochmann2013}, resonantly enhanced electro-optics ~\cite{Rueda2016,Fan2018}, solid-state emitter ensembles~\cite{Williamson2014,OBrien2014,Zhong2017}, and trapped atoms~\cite{Hafezi2012,Hogan2012,Pritchard2014,Kiffner2016,Gard2017,Han2018} to achieve strong nonlinearities. However, several of these approaches require integrating superconductors in close proximity to high-intensity optical fields~\cite{Higginbotham2018,Fan2018}. This can have deleterious effects on the microwave resonator Q-factors, and in the case of nanomechanical resonators it introduces noise photons into the conversion process. 

\begin{figure}[ht]
\centering
\includegraphics[width=8.6cm]{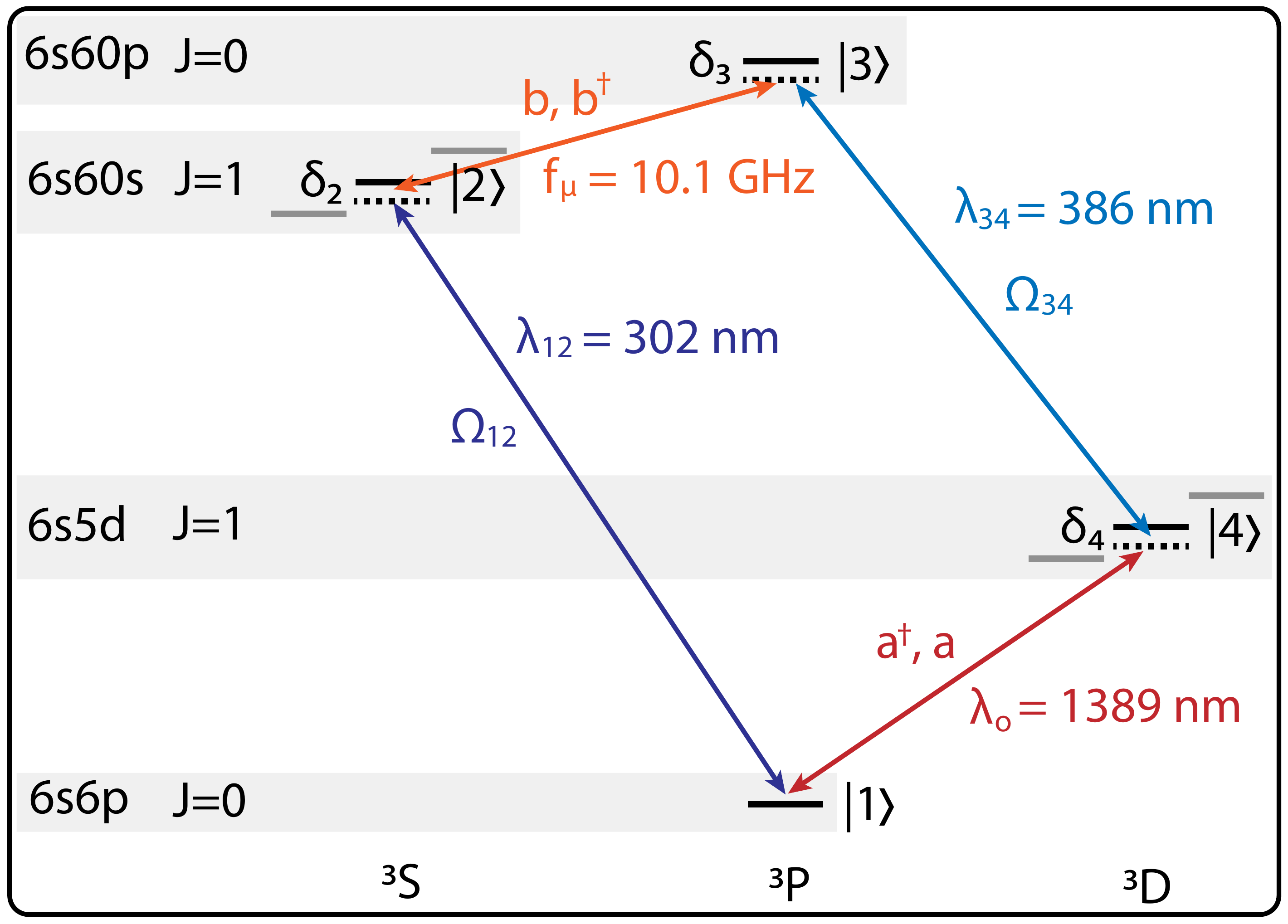}
\caption{\textbf{The conversion loop}. The atom is initialized in the $6s6p$ $^3$P$_0$ `clock' state ($|1\rangle$). This state is in a loop of three other states which include the $6s60s$ $^3$S$_1$ Rydberg state ($|2\rangle$), the $6s60p$ $^3$P$_0$ Rydberg state ($|3\rangle$), and the $6s5d$ $^3$D$_1$ state ($|4\rangle$). The two Rydberg states $|2\rangle$ and $|3\rangle$ are spaced by 10.1 GHz, and strongly coupled by an electric dipole moment. The states $|4\rangle$ and $|1\rangle$ are separated by a telecom-band photon with wavelength $\lambda_\text{o}=1389$ nm. We consider continuous wave operation in which the coupling fields are detuned from the intermediate states. The population primarily resides in $|1\rangle$.}
\label{fig:Figure1}
\end{figure}

In the neutral atom approach, coupling to microwave photons is most notably mediated by transitions between two highly-excited Rydberg states since they have large transition electric dipole moments~\cite{Petrosyan2009,Hogan2012,Pritchard2014,Kiffner2016,Gard2017,Han2018}. The desired microwave frequency can be selected simply by choosing the principle quantum number $n$ of the Rydberg levels appropriately. All such efforts focus on alkali atoms in which optical transitions in the telecommunication window are only accessible via complex schemes involving six or seven internal states, many of which are short-lived~\cite{Pritchard2014,Kiffner2016,Gard2017,Han2018}. Significantly, the complexity associated with six- and seven-wave-mixing makes continuous-wave (CW) operation with significant detuning from intermediate states challenging, and most efforts have focused on resonant pulsed operation. Indeed, some efforts even use pulse design with optimal control techniques~\cite{Gard2017}. Moreover, the nuclear spins of the alkalis give rise to hyperfine structure in the Rydberg manifold, resulting in undesirable additional off-resonant couplings.

Here, we propose detuned, continuous-wave (CW) four-wave-mixing~\cite{Brekke2008,Willis2009,Becerra2010} to achieve CW microwave-to-optical transduction in an ensemble of ytterbium (Yb) atoms, which have a strong transition in the telecommunication E-band at 1389 nm from a long-lived metastable state~\cite{Covey2019} (see Fig.~\ref{fig:Figure1}). Specifically, we consider $N=150$ $^{174}$Yb atoms with zero nuclear spin in a cigar-shaped geometry~\cite{Kiffner2016}. The ensemble is placed inside a copper microwave resonator with $Q_\mu=4.5\times10^3$ at resonance frequency $f_\mu=10.1$ GHz. While the large optical density along this axis gives rise to a collective enhancement of the optical interaction, a small optical cavity linewidth ($\approx100$ kHz) is required for good impedance matching, and thus a resonator with $\mathcal{F}=1.4\times10^4$ is needed. The use of an ensemble results in a large cooperativity on the optical transition. We demonstrate that bandwidths of $\approx100$ kHz are possible in detuned, continuous-wave (CW) operation.

\section{Overview of the conversion process}
We begin with an overview of the parametric conversion cycle, including properties of the levels involved, and the required phase matching conditions. The four-level cycle operates in a diamond configuration whose lowest level is the `clock' state $^3$P$_0$ of lifetime $>20$ s, which can be accessed from the ground state $^1$S$_0$ by either direct excitation~\cite{Barber2006,Ludlow2015} or by multi-photon excitation via $^3$P$_1$~\cite{Barker2016} (see Ref.~\cite{Covey2019} for a more complete level diagram and Appendix II.A for analysis of the branching ratios to the $^3$P$_J$ manifold). The diamond configuration of the transducer cycle is shown in Fig.~\ref{fig:Figure1}, where the four states of the cycle are $|\{1,2,3,4\}\rangle\equiv|\{6s6p~^3\text{P}_0,6s60s~^3\text{S}_1,6s60p~^3\text{P}_0,6s5d~^3\text{D}_1\}\rangle$.

The dipole matrix element (DME) of the $^3$D$_1$$\rightarrow$$^3$P$_0$ transition is $d_\text{o}=1.63$ e-a$_0$~\cite{Covey2019}. The DMEs for the two optical transitions to the Rydberg states $d_{12}$ and $d_{34}$ can be estimated using the Coulomb wavefunction approach as described in Appendix II. They are determined to be $d_{12}=2.0\times10^{-3}$ e-a$_0$ and $d_{34}=3.7\times10^{-4}$ e-a$_0$. The Rydberg-Rydberg transition dipole matrix element for $n=60$ is $d_{23}=d_\mu=1436$ e-a$_0$. 

\begin{figure}[ht]
\centering
\includegraphics[width=8.6cm]{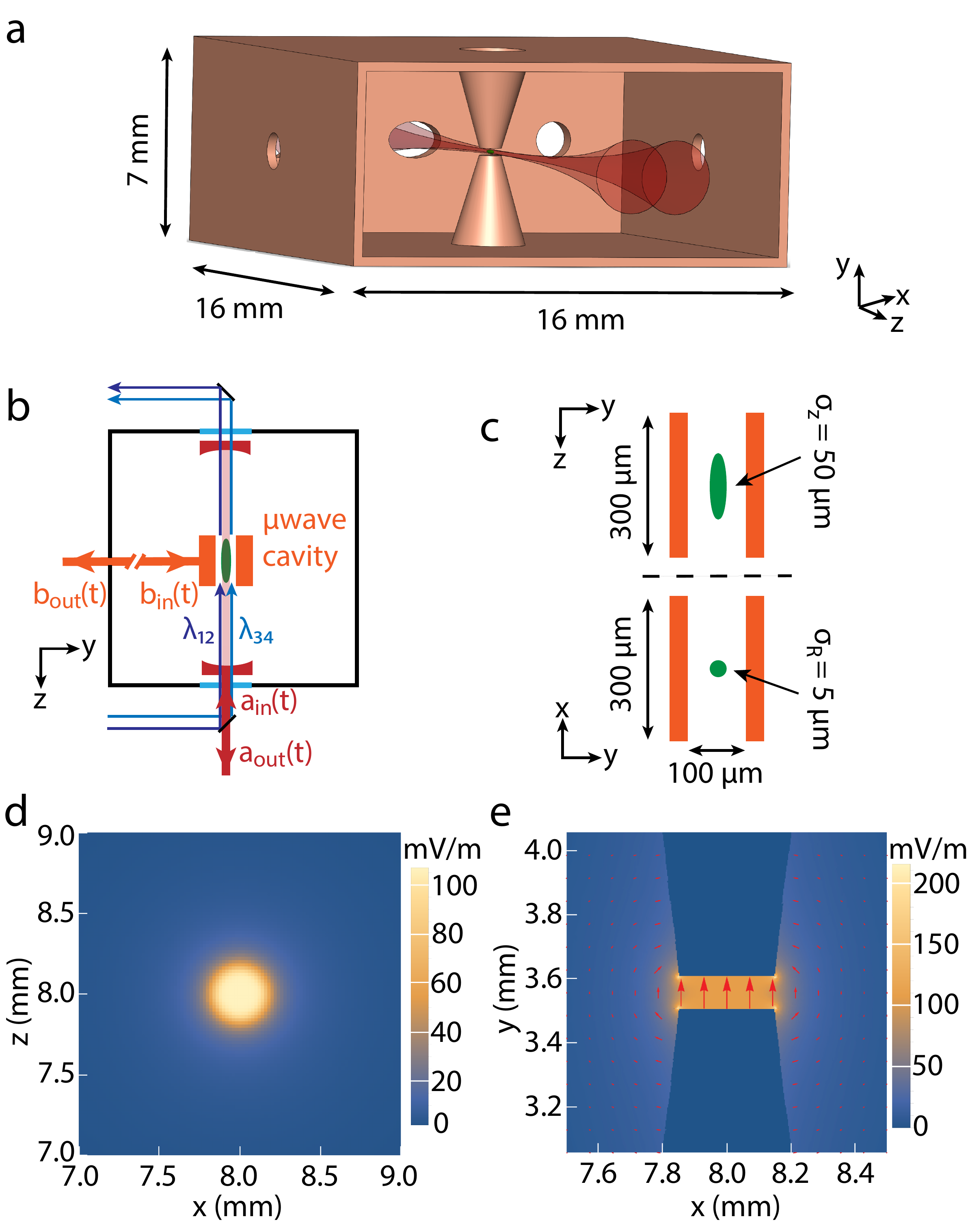}
\caption{\textbf{The transduction circuit in a cryogenic environment} (a) The copper microwave resonator with round capacitor plates on conically-tapered pillars (not to scale). The red beams show the optical dipole trapping of the cigar-shaped cloud (green). (b) The cryogenic environment with transduction circuit inside. The black box shows the cryostat surfaces, and the blue shows the viewports for the optical fields, which can be combined using a dichroic mirror. A traveling microwave photon in mode $b_\text{in}(t)$  from an external superconducting circuit enters the microwave resonator (orange) containing the atomic cloud (green ellipse). The UV (purple) and blue (blue) driving fields, and the telecommunication-wavelength cavity mode (red) are coaxial to obtain phase matching in the atomic medium. (c) A zoom-in of the microwave resonator plates and the atomic ensemble, shown on the $yz$-plane and the $xy$-plane. Simulated root-mean-square (RMS) electric field magnitude per single photon shown on the yz-plane (d) and xy-plane (e). The vector field between the resonator plates is shown in (e). The field amplitude and orientation is uniform across the atomic cloud.}
\label{fig:Figure2}
\end{figure}

We now consider the phase matching conditions in order to gain insight on the required propagation axes of the four fields. The phase matching conditions require $\omega_{12}+\omega_{23}=\omega_{34}+\omega_{41}$ and $\vec{k}_{12}+\vec{k}_{23}=\vec{k}_{34}+\vec{k}_{41}$, where $\omega$ is the angular transition frequency and $\vec{k}$ is the wavevector of the photon in the atomic medium. The frequencies are $\{\omega_{12},\omega_{23},\omega_{34},\omega_{41}\}/(2\pi)=\{990,10.1\times10^{-3},780,216\}$ THz~\cite{VidolovaAngelova1981}. Assuming the detunings $\delta_i$ are large compared to the Rabi frequencies $\Omega_i$ and the linewidths $\gamma_i$, the wavenumbers are $\{k_{12},k_{23},k_{34},k_{41}\}/(2\pi)=\{33100,0.34,25900,7200\}$ cm$^{-1}$. Accordingly, $\omega_{23}$ and $k_{23}$ are negligible, which allows $\theta_{12}\approx\theta_{34}\approx\theta_{41}=0$ degrees with respect to the $z$-axis, and $\theta_{23}=90$ degrees (i.e. along the $y$-axis), where $\vec{k}\cdot\hat{r}=k~\text{cos}(\theta)$ and $\hat{r}$ points along the $z$-axis and long-axis of the ensemble (see Fig.~\ref{fig:Figure2}a). This configuration allows for a crossed-cavity geometry of the microwave and optical resonators.

\section{Formulation via adiabatic elimination}
We focus on the case of large detunings, $|\delta_i|>|\Omega_i|$ for $i\in\{2,4\}$ (note that excitation to $|3\rangle$ is a two-photon process, and the requisite conditions are analyzed below), for which we can adiabatically eliminate the intermediate states $|2\rangle$, $|3\rangle$, and $|4\rangle$~\cite{Brion2007,Williamson2014}. Such an approach is analogous to work with solid state spin ensembles~\cite{Williamson2014,Zhong2017}, and it allows us to apply an input-output formalism~\cite{Collett1984,Gardiner1985,Gardiner1993} for the microwave and optical fields. 

We begin by considering the effective Hamiltonian describing the coupling between the microwave and optical modes. The relevant terms have a linear coupling, which upon the adiabatic elimination of the excited states has the form~\cite{Brion2007,Williamson2014}:
\begin{equation}
H_\text{eff}/\hbar=(S a^\dagger b + S^*b^\dagger a),
\end{equation}
where $a^\dagger$ ($a$) is the creation (annihilation) operator for the optical field, and $b^\dagger$ ($b$) is the creation (annihilation) operator for the microwave field, and
\begin{equation}
S=\frac{\tilde{\Omega}_{12}\Omega_{34}^*g_\mu \tilde{g}_\text{o}^*}{\delta_2\delta_3\delta_4}F.
\end{equation}

$S$ describes the coupling strength between the microwave and optical modes. $\tilde{\Omega}_{12}=\sqrt{N}\Omega_{12}$ is the collectively-enhanced Rabi frequency of the $|1\rangle-|2\rangle$ transition for $N=150$ atoms, $g_\mu$ is the single-atom coherent coupling rate to the microwave field for the $|2\rangle-|3\rangle$ transition, $\Omega_{34}$ is the single-atom Rabi frequency of the $|3\rangle-|4\rangle$ transition, and $\tilde{g}_\text{o}=\sqrt{N}g_\text{o}$ is the collectively-enhanced coherent coupling rate to the optical field of the $|4\rangle-|1\rangle$ transition. See Appendix I for a discussion of the Hamiltonian and the collective states. Note that \textit{a priori} each quantity is atom-dependent, for which we could sum over the atoms in the ensemble. As shown in Appendix IV, inhomogeneous broadening is negligible, and thus we consider collectively-enhanced quantities. The detunings $\delta_i$ are defined in Fig.~\ref{fig:Figure1}.

The factor $F$ contains the effects of imperfect mode overlap as an effective filling factor~\cite{Williamson2014}:
\begin{equation}
F=\frac{1}{V_e}|\int_{V_e}\chi_{12}(\mathbf{r})\chi_\mu(\mathbf{r})\chi_{34}(\mathbf{r})\chi_\text{o}(\mathbf{r})d^3\mathbf{r}|,
\end{equation}
where $V_e$ is the volume of the atomic ensemble. The $\chi_\mu(\mathbf{r})$, $\chi_{12}(\mathbf{r})$,  $\chi_\text{o}(\mathbf{r})$, and $\chi_{34}(\mathbf{r})$ are the mode functions for the microwave, and three optical modes, respectively.  They are defined as the slowly varying electric field amplitudes of the optical and microwaves fields, and the phase matching is satisfied as discussed in Sec. II. Here, the field distribution is defined where the maximal value at the center of the atomic cloud is set to unity. Hence, the range of this factor is between zero and one. The uniformity of the microwave and the optical fields over the atomic cloud results in a good overlap with $F\approx0.9$. 

Now we apply the input-output formalism~\cite{Collett1984,Gardiner1985,Gardiner1993} to arrive at the relations between the optical (microwave) resonator input $a_\text{in}(t)$ ($b_\text{in}(t)$) and output modes $a_\text{out}(t)$ ($b_\text{out}(t)$). We immediately apply a Fourier transform to arrive at:
\begin{multline}
\tilde{a}_\text{out}(\omega)=\frac{iS\sqrt{\kappa_\text{e}^\text{o}\kappa_\text{e}^\mu}}{\tilde{\kappa}_\text{o}\tilde{\kappa}_\mu(1+\tilde{C}_{\text{o}\mu})}\tilde{b}_\text{in}(\omega) \\
+\bigg(1-\frac{\kappa_\text{e}^\text{o}}{(1+\tilde{C}_{\text{o}\mu})\tilde{\kappa}_\text{o}}\bigg)\tilde{a}_\text{in}(\omega),
\end{multline} 
and similarly for $\tilde{b}_\text{out}$:
\begin{multline}
\tilde{b}_\text{out}(\omega)=\bigg(1-\frac{\kappa_\text{e}^\mu}{(1+\tilde{C}_{\text{o}\mu})\tilde{\kappa}_\mu}\bigg)\tilde{b}_\text{in}(\omega)
\\
+\frac{iS\sqrt{\kappa_\text{e}^\text{o}\kappa_\text{e}^\mu}}{\tilde{\kappa}_\text{o}\tilde{\kappa}_\mu(1+\tilde{C}_{\text{o}\mu})}\tilde{a}_\text{in}(\omega).
\end{multline}

\begin{figure}[ht]
\centering
\includegraphics[width=8.6cm]{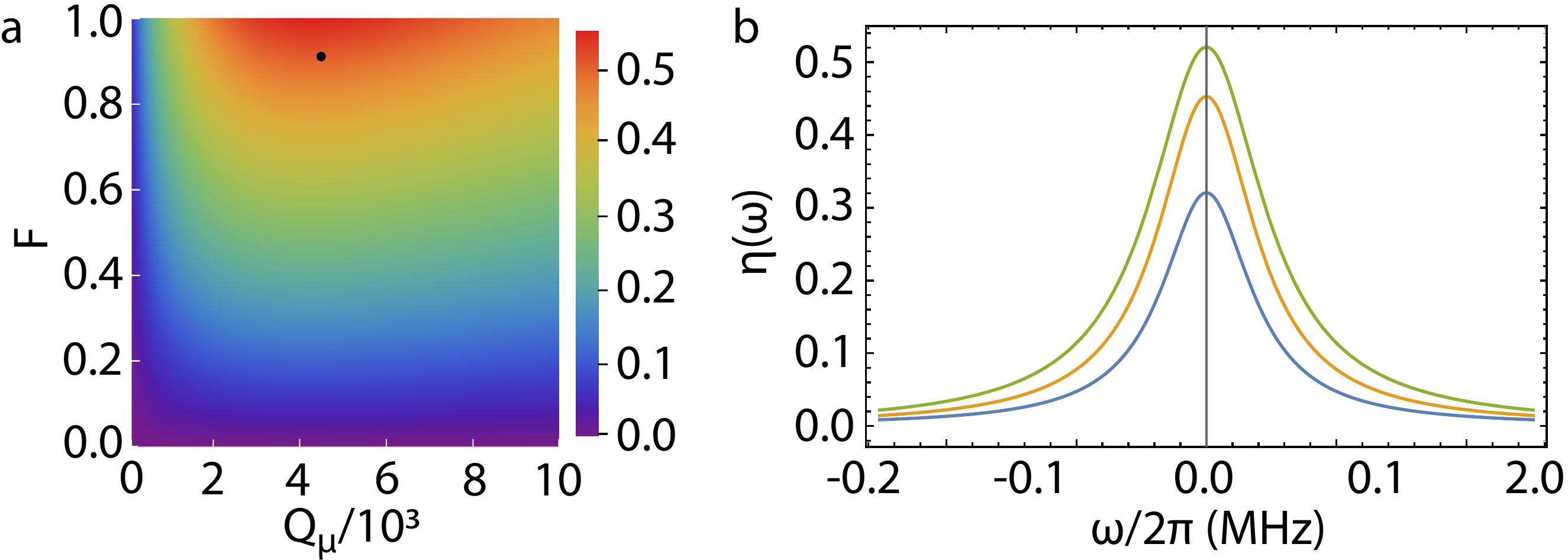}
\caption{\textbf{Impedance matching and bandwidth} (a) The conversion efficiency $\eta(\omega=0)$ as a color plot ranging from 0.0 to 0.6. It is shown as a function of the mode filling fraction $F$ on the vertical axis and the Q-factor of the microwave resonator on the horizontal axis. The thick black dot corresponds to the proposed values in this work. (b) The conversion efficiency $\eta(\omega)$ as a function of frequency $\omega$. The three colors correspond to $Q_\mu=1\times10^3$ (blue, bottom), $Q_\mu=2\times10^3$ (orange, middle), and $Q_\mu=4.5\times10^3$ (green, top). We focus on the case of $Q_\mu=4.5\times10^3$ throughout (which corresponds to the black dot in (a)), for which the FWHM bandwidth is $\approx100$ kHz.}
\label{fig:Figure3}
\end{figure}

$\tilde{\kappa}_\text{o}=(\kappa_\text{e}^\text{o}+\kappa_\text{i}^\text{o})/2-i\omega$, $\tilde{\kappa}_\mu=(\kappa_\text{e}^\mu+\kappa_\text{i}^\mu)/2-i\omega$, and $\tilde{C}_{\text{o}\mu}=|S|^2/\tilde{\kappa}_\text{o}\tilde{\kappa}_\mu$, which is the effective cooperativity for the full system. $\kappa_\text{e}^\text{o}$ and $\kappa_\text{e}^\mu$ are the external bandwidths for the optical and microwave resonator, respectively, and $\kappa_\text{i}^\text{o}$ and $\kappa_\text{i}^\mu$ are the intrinsic bandwidths limited by the materials of the cavities. The Fourier frequency $\omega=0$ corresponds to the resonance condition, and hence we can define $\omega=\omega_{34}+\omega_\text{o}-\omega_{12}-\omega_\mu$. Note that this formalism generically applies to two coupled resonators, and the details of the adiabatically-eliminated excited states are all contained in the coupling strength parameter $S$. 

In the limit that the intrinsic linewidths $\kappa_\text{i}$ are taken to zero, equations (4) and (5) reduces to the following spectral relation obtained in Ref.~\cite{Williamson2014}:
\begin{multline}
\tilde{a}_\text{out}(\omega)=\frac{4iS\sqrt{\kappa_\text{o}\kappa_\mu}}{4|S|^2+(\kappa_\text{o}-2i\omega)(\kappa_\mu-2i\omega)}\tilde{b}_\text{in}(\omega) \\
+\frac{4|S|^2-(\kappa_\text{o}+2i\omega)(\kappa_\mu-2i\omega)}{4|S|^2+(\kappa_\text{o}-2i\omega)(\kappa_\mu-2i\omega)}\tilde{a}_\text{in}(\omega),
\end{multline}
and
\begin{multline}
\tilde{b}_\text{out}(\omega)=\frac{4|S|^2-(\kappa_\text{o}+2i\omega)(\kappa_\mu-2i\omega)}{4|S|^2+(\kappa_\text{o}-2i\omega)(\kappa_\mu-2i\omega)}\tilde{b}_\text{in}(\omega)
\\
+\frac{4iS\sqrt{\kappa_\text{o}\kappa_\mu}}{4|S|^2+(\kappa_\text{o}-2i\omega)(\kappa_\mu-2i\omega)}\tilde{a}_\text{in}(\omega).
\end{multline}

The first term in Eq. (6) and the second term in Eq. (7) describe the photon conversion amplitude between the microwave and optical fields, and the square of those terms give the conversion efficiency $\eta(\omega)$~\cite{Williamson2014}. Accordingly, we obtain a impedance matching condition required to reach $\eta=1$ when $\omega=0$: $4|S|^2=\kappa_\text{o}\kappa_\mu$~\cite{Collett1984,Gardiner1985,Gardiner1993,Williamson2014}. This relation provides a concrete approach to designing the transduction circuit and quantifying the bandwidth and efficiency.

For the microwave and optical cavities described below and shown in Fig.~\ref{fig:Figure2}, the single-atom coupling rates are $g_\mu = 2\pi\times1.4$ MHz and $g_\text{o}=2\pi\times130$ kHz, respectively. The collective coupling rate for the latter is $\tilde{g}_\text{o}=2\pi\times1.5$ MHz with $N=150$ atoms. For convenience, we choose $g_\mu\simeq\tilde{\Omega}_{12}=\Omega_{34}=\tilde{g}_\text{o}$. Naturally, we must choose the detunings to maintain adiabaticity and to limit the off-resonant scattering rates to well below the conversion rate. We take $\delta_2=2\pi\times4.5$ MHz, $\delta_3=2\pi\times1.5$ MHz, and $\delta_4=2\pi\times4.5$ MHz. 
The role of small populations in the excited states and spontaneous emission are analyzed in Appendix III.

For these values and the other parameters listed below, we find $S=2\pi\times140$ kHz. We focus on $Q_\mu=4.5\times10^3$ and $Q_\text{o}=4\times10^9$. Figure~\ref{fig:Figure3}a shows $\eta(\omega=0)$ as a density plot of the filling factor $F$ and the microwave resonator $Q_\mu$. The proposed conditions are shown as the thick black dot, for which $\eta(\omega=0)\approx0.5$. The bandwidth $\eta(\omega)$ is shown in Fig.~\ref{fig:Figure3}b for values of $Q_\mu$. For $Q_\mu=4.5\times10^3$, the FWHM bandwidth is $\tilde{\Gamma}_{\text{o}\mu}/2\pi\approx100$ kHz, which approaches typical bandwidths reached in superconducting quantum circuits~\cite{Devoret2013,Barends2014}. 

\section{Parameters}
We now turn to the design of the experiment to obtain all the relevant quantities. Thermal excitations of the microwave mode will result in additional noise photons during transduction. To reduce the thermal photon noise level the transducer needs to be operated inside a dilution refrigerator where the thermal occupancy can be reduced below $n_{th}\leq0.01$ at $100\,$mK. However, we note that in an efficient converter, the microwave cavity photons decay primarily to the optical and microwave output fields at lower temperatures instead of the phonon temperature of the resonator. This effect results in a reduced thermal noise in the transduction and has been utilized for proposals using high-temperature quantum communication with microwave photons~\cite{Xiang2017}. 

In addition to the thermal microwave photons, the optical trapping and driving fields will result in additional heating (see Appendix V for further discussion). It is therefore important to design the system to minimize the optical power and light scattering using anti-reflection coated radiation shield windows. Initial construction and testing of this system at $4$ K could therefore be enlightening before the single-photon level is approached.

\subsection{The microwave resonator}
Since the Rydberg-Rydberg microwave transition is an electric dipole transition, we employ a microwave resonator where the two conically-tapered, extruded, disc-shaped plates in the box confine the electric field energy to a small mode volume (see Fig.~\ref{fig:Figure2}a). The geometry of the resonator is shown in Fig.~\ref{fig:Figure2}a \& 2c. The plates are separated by 100 $\mu$m, and have a radius of $150$ $\mu$m. This geometry was chosen to match the size of the atomic ensemble, though the mode volume is larger than the profile of the atomic ensemble to ensure homogeneity across the cloud. DC Stark shifts of the Rydberg atoms are expected to be negligible in our system at cryogenic temperatures~\cite{Gard2017,Booth2018}. The copper box has a volume of $16\times16\times7$ mm$^3$, and can have many holes for optical beam paths, as in Fig.~\ref{fig:Figure2}a.

We simulate the mode distribution of the cavity using the COMSOL finite-element-method solver (see Fig.~\ref{fig:Figure2}c), and obtain a root-mean-square (RMS) electric field amplitude of $E^{RMS}_\mu\approx110$ mV/m per photon with excellent uniformity in magnitude and direction over the volume of the cloud. The vacuum Rabi frequency (i.e. corresponding coherent coupling rate) $g_\mu$ is given in terms of the electric field as $g_\mu=\frac{d_\mu\cdot \sqrt{2}E^{RMS}_\mu}{2\hbar}$, from which we obtain $g_\mu/2\pi\approx1.4$ MHz.

We choose a moderate quality factor of $Q_\text{e}^\mu=4.5\times10^3$, which is small compared to the intrinsic quality factor for copper resonators of $Q_\text{i}^\mu\approx10^4$ (see e.g. Refs.~\cite{Lamoreaux2013,Probst2014,Brubaker2017}). Hence, it is possible to maintain high conversion efficiencies. The corresponding external linewidth for our resonator operating at $10.1$ GHz is $\kappa_\text{e}^\mu\approx2\pi\times2.2$ MHz.

The use of copper instead of a high-Q superconducting resonator relaxes the constraints on the applied magnetic fields. Though we operate the conversion cycle with only a small quantization magnetic field, laser cooling and initialization may require fields on the order of 10 G. Since magnetic fields are known to decrease the $Q$ of superconducting resonators, we choose copper for simplicity. The effects of the incident optical power on the resonator are discussed in Appendix V.

\subsection{The optical resonator and pump modes}
We assume the optical dipole trapping beam, the pump fields $\tilde{\Omega}_{12}$ and $\Omega_{34}$, and the optical field $\tilde{g}_\text{o}$ all propagate in the $z$-direction and have approximately the same mode profile. The optical field $\tilde{g}_\text{o}$ is assumed to have a $1/e^2$ waist radius of $37$ $\mu$m in order to roughly match the radial size of the atomic ensemble (see Fig.~\ref{fig:Figure2} and subsection IV.c) while still facilitating stable optical cavity operation. The axial profiles of all optical modes are determined by their respective Rayleigh lengths dictated by Gaussian optics. 

We propose the use of a moderate-finesse optical resonator with a finesse $\mathcal{F}=1.4\times10^4$. We assume asymmetric reflectivities of the cavity mirrors such that the photon is preferentially extracted on one side of the cavity, as in Fig.~\ref{fig:Figure2}b. Such a resonator allows the possibility of coupling in all the optical fields while still providing a sufficient optical quality factor $Q_\text{e}^\text{o}$. Since the finesse is related to the mirror reflectivities $r_1$ and $r_2$ and loss per pass $\xi$ by $\mathcal{F}(r_1,r_2,\xi)=\pi\sqrt[4]{r_1r_2(1-\xi)}/(1-\sqrt{r_1r_2(1-\xi)})$, we see that a finesse of $\mathcal{F}=1.4\times10^4$ requires reflectivities of $r_1=0.99998$ and $r_2=0.9996$ with a loss per pass of $\xi=0.00001$. This allows us to use dichroic coatings to inject the 302 and 386 nm pump beams and the 760 nm trapping beam along the cavity axis into the optical resonator via the mirror with reflectivity $r_2$ (see Fig.~\ref{fig:Figure2}b).

Now we can estimate $Q_\text{o}$ based on the proposed optical resonator. We assume a cavity length of $L=20$ cm. In order to achieve a mode waist of $w_0^\text{o}=37$ $\mu$m the radius of curvature of the resonator mirrors must be $R_c=10.01$ cm. This value is near the stability edge of $L/2$, but expected to be robust under the proposed conditions. The mode volume of the resonator is $V_\text{o}=\pi/4(w_0^\text{o})^2L=0.2$ mm$^3$. The corresponding free spectral range of this cavity is $\text{FSR}=2\pi\times750$ MHz, for which the linewidth $\kappa_\text{e}^\text{o}=\text{FSR}/\mathcal{F}=2\pi\times54$ kHz. Accordingly, the quality factor of this resonator is $Q_\text{e}^\text{o}=\omega_{o}/\kappa_\text{e}^\text{o}=4\times10^9$. We assume an intrinsic optical quality factor of $Q_\text{i}^\text{o}=10Q_\text{e}^\text{o}$, which will be limited only by the optical coating required to insert the UV pump fields. 

We calculate the coherent coupling rate using $V_\text{o}$, and we find $g_\text{o}/2\pi\approx130$ kHz. This quantity is collectively enhanced by the ensemble, and the collective coupling rate is $\tilde{g}_\text{o}/2\pi\approx1.5$ MHz. Accordingly, the system is in the strong coupling regime and the cavity induced atomic decay rate $\Gamma_{1\text{D}}= (\tilde{g}_\text{o}/\delta_4)^2 \kappa_e$ is much less than the free space spontaneous emission rate $\Gamma_4$.

We assume the optical pump modes $\tilde{\Omega}_{12}$ and $\Omega_{34}$ have ($1/e^2$) waists of $25$ $\mu$m in order to be homogeneous over the ($1/e$) $\sigma_\text{R}=5$ $\mu$m radial extent of the atomic ensemble. The Rabi frequencies used above were chosen to be $\tilde{\Omega}_{12}=\Omega_{34}=2\pi\times1.5$ MHz. Based on the DMEs of these transitions, we estimate that powers of $P_{12}\approx120$ $\mu$W and $P_{34}\approx500$ mW are required. While reaching $\lambda_{12}$ will likely require two nonlinear stages such as in fourth harmonic generation, $\lambda_{34}$ can be reached via second harmonic generation of a 772 nm light from a Titanium-Saphhire laser. Such powers can be achieved with commercial or home-made lasers.

\subsection{Optical trapping}
The atomic ensemble is trapped in two beams crossed at a shallow angle along the x-axis propagating through the microwave resonator (along the $z$-direction), which naturally gives rise to the extended cigar-shaped geometry. We propose an optical dipole trapping wavelength of 760 nm, which traps both the $^1$S$_0$ and $^3$P$_0$ states of Yb in a ``magic'' condition~\cite{Ludlow2015}. We consider $N=150$ atoms in a cloud of sizes ($1/e$ radius) $\sigma_\text{R}\approx5$ $\mu$m and $\sigma_\text{Z}\approx50$ $\mu$m along the radial and axial directions, respectively. This results in a modest density of $\rho\approx2\times10^{10}$ cm$^{-3}$, which can be reached directly after narrow-line laser cooling.

The typical temperature of a Yb magneto-optical trap operating on the $^1$S$_0$$\rightarrow$$^3$P$_1$ transition is $T=5-10$ $\mu$K~\cite{Yamamoto2016}, and thus a dipole trap of depth $\approx20$ $\mu$K is sufficient. We assume a trapping beam waist ($1/e^2$ radius) of $\approx25$ $\mu$m, and we assume two beams crossed at a shallow angle of $\approx1$ degrees such that the trap aspect ratio matches the desired aspect ratio of the cloud. These beams require a total power of $\approx0.5$ W. As such, negligible power is incident on the microwave resonator. The total cooling power of $\approx 1.5\,$mW is available on the cold plate at 120 mK in state-of-the art commercial dilution fridge systems. However, significantly less power should be incident on the copper resonator to avoid optically induced microwave noise photons. See Appendix V for a discussion of thermal management of the optical power. We leave a more detailed analysis of the cooling and trapping scheme for further investigation.

\section{Conclusion and Outlook}
We propose an Yb ensemble as a viable approach to MOC, and we show that bandwidths of $\approx100$ kHz are possible with efficiencies of $\approx50\%$. While this is a challenging experimental platform, we believe its capability and simplicity are favorable compared to the current efforts with alkali atoms. Further, there are several on-going experimental efforts with cryogenic alkaline earth atoms for applications in precision metrology~\cite{Ushijima2015,Ludlow2015}. Moreover, the wavelength of the optical field used here allows the MOC circuit to be interfaced with quantum repeater stations based on the same optical transition in Yb atoms, as proposed recently~\cite{Covey2019}, and could facilitate a long-distance quantum network of superconducting quantum computers.

We note that our approach could be modified in several ways to increase transduction bandwidth and relax thermal requirements. The use of a strong, resonant drive $\Omega_{12}$ would enable collective enhancement on both the optical and microwave quantum fields. Noise and broadening analysis in such a scheme is beyond the scope of this work. A wide range of microwave frequencies could be chosen by targeting different Rydberg states. For instance, if we instead choose a frequency of 100 GHz, the corresponding temperature is $\approx5$ K where thermal management is greatly simplified. With continued advances in quantum frequency conversion with nonlinear microwave circuits~\cite{Roch2012,Abdo2013,Inomata2014}, single photons from superconducting circuits operating in the $\approx5-10$ GHz could be upconverted to a higher frequency band for MOC. 

Finally, our system could be extended to $^{171}$Yb, an isotope with nuclear spin $I=1/2$, to introduce a long-lived quantum memory. This could enable entanglement between a microwave photon and the nuclear spin by the same mechanism as recently proposed for an optical photon~\cite{Covey2019}, and could provide a platform for integrating a quantum memory with superconducting quantum circuits. Such a possibility showcases the potential of using cold alkaline-earth atoms, in which nuclear coherence times far exceed that of solid-state platforms.

\section*{Acknowledgements}
We acknowledge Oskar Painter, Manuel Endres, Jonathan Simon, Jevon Longdell and Jono Everts for helpful discussions, and Ivaylo Madjarov for stimulating discussions about the use of master equations. JPC acknowledges support from the Caltech PMA Division for post-doctoral fellowship funding, and AS acknowledges support from the Caltech IQIM for post-doctoral fellowship funding. MS was supported by Army Research Office, Contract No. W911NF-16-1-0133, and by the US Army Research Laboratory Center for Distributed Quantum Information through Cooperative Agreement No. W911NF-15-2-0061. 

\setcounter{section}{0}  
\renewcommand{\thesection}{APPENDIX~\Roman{section}}

\section{Derivation of the Hamiltonian and collective states}
The system Hamiltonian described in Fig.~\ref{fig:Figure1} is given by:
\begin{multline}
    H_\text{Sys}/\hbar=\sum_k (\delta_{2,k}\sigma_{22,k}+\delta_{3,k}\sigma_{33,k}+\delta_{4,k}\sigma_{44,k}) \\
    +\sum_k(\Omega_{12,k}\sigma_{12,k}+\text{H.c.})+\sum_k(g_{\mu,k}\sigma_{23,k}b+\text{H.c.})
    \\
    +\sum_k(\Omega_{34,k}\sigma_{34,k}+\text{H.c.})+\sum_k(g_{\text{o},k}\sigma_{41,k}a+\text{H.c.}),
\end{multline}
where the index $k$ runs over atoms in the ensemble, and $\sigma_{ij,k}=|i\rangle\langle j|$ are matrix elements in the four-state Hilbert space.

We work with large detunings, where $|\delta_{\text{o},k}|>|g_{\text{o},k}|$, $|\delta_{\mu,k}|>|g_{\mu,k}|$, and $|\delta_{\text{o},k}\delta_{\mu,k}|>|\Omega_{12,k}\Omega_{34,k}|$. Note that $\delta_{\text{o},k}=\delta_{4,k}$, and $\delta_{\mu,k}=\delta_{3,k}-\delta_{2,k}$. Hence, we can adiabatically eliminate the excited states~\cite{Brion2007} to arrive at the following effective Hamiltonian:
\begin{multline}
    H_\text{eff}/\hbar=\sum_k \bigg(-\frac{|g_\text{o}|^2}{\delta_{\text{o},k}}a^\dagger a -\frac{|g_\mu|^2}{\delta_{\mu,k}}b^\dagger b
    \\
    +\frac{\Omega_{12,k}\Omega_{34,k}^*g_{\mu,k}g_{\text{o},k}^*}{\delta_{2,k}\delta_{3,k}\delta_{4,k}}a^\dagger b+\frac{\Omega_{12,k}^*\Omega_{34,k}g_{\mu,k}^*g_{\text{o},k}}{\delta_{2,k}\delta_{3,k}\delta_{4,k}}b^\dagger a\bigg).
\end{multline}
Note that a.c. Stark shifts from the detuned drive fields have been neglected based on the assumption of large-detunings, and based on the analysis of inhomogeneous broadening in Appendix IV. Likewise, we neglect inhomogeneous effects. Since the phase matching condition is satisfied, each atomic contribution adds constructively and  and we can  replace the sums over atoms with a factor of $N$. 

The first two terms are due to off-resonant atoms pulling the resonant frequencies of the two cavities. The latter two terms are a linear coupling between the two modes with strength S, as described in the main text. 

We now analyze the collectively-excited Dicke states in the ensemble which result from the above Hamiltonian. We define the initial state in which the ensemble is polarized as $|\tilde{1}\rangle=|1\rangle^{\otimes N}$. %, where $|0\rangle$ denotes no excitations in the Fock basis. 
In the limit of one excitation distributed amongst the atoms in the ensemble, the normalized collective states $|\tilde{2}\rangle$, $|\tilde{3}\rangle$, and $|\tilde{4}\rangle$ are given accordingly by  $|\tilde{2}\rangle=1/\sqrt{N}\sum_k\sigma_{21,k}|\tilde{1}\rangle$, $|\tilde{3}\rangle=1/\sqrt{N}\sum_k\sigma_{31,k}|\tilde{1}\rangle$, and $|\tilde{4}\rangle=1/\sqrt{N}\sum_k\sigma_{41,k}|\tilde{1}\rangle$, where $\sigma_{n1,k}$ excites atom $k$ from state $|1\rangle$ to state $|n\rangle$. We assumed phase matching condition and omitted the relative atomic phases for clarity.

% where $|1_\alpha\rangle$ denotes excitations in the Fock basis in state $|\alpha\rangle$ for $\alpha=2,~3,~4$, and $P_i$ ranges over all operators that permute the $N$ qubits in the $N$ possible distinct ways. This can also be expressed as $|\tilde{2}\rangle=1/\sqrt{N}\sum_k\sigma_{21,k}|\tilde{1}\rangle$, $|\tilde{3}\rangle=1/\sqrt{N}\sum_k\sigma_{31,k}|\tilde{1}\rangle$, and $|\tilde{4}\rangle=1/\sqrt{N}\sum_k\sigma_{41,k}|\tilde{1}\rangle$.

To analyze the effective Rabi frequencies between collective states $\tilde{\Omega}_{ij}$, we consider $\langle\tilde{i}|H_{Sys}|\tilde{j}\rangle$ for $j-i=1$ and $i=1,~2,~3$. First, we consider $\tilde{\Omega}_{12}$:
\begin{multline}
    \tilde{\Omega}_{12}=\bigg(\frac{1}{\sqrt{N}}\sum_i\langle\tilde{1}|\sigma_{12,i}^*\bigg)\bigg(\sum_j\Omega_{12,j}\sigma_{12,j}\bigg)|\tilde{1}\rangle
    \\
    =\frac{1}{\sqrt{N}}\sum_{i=j}\Omega_{21,i}=\sqrt{N}\Omega_{21,k},
\end{multline}
where the coupling for each atom is assumed to be identical (see Appendix IV). Note that since the ensemble is initially polarized in $|\tilde{1}\rangle$, this collective state is fundamentally different from $|\tilde{2}\rangle$, $|\tilde{3}\rangle$, $|\tilde{4}\rangle$ within the single-excitation Dicke manifold. Accordingly, $\tilde{\Omega}_{41}=\tilde{g}_\text{o}$ will be similar to $\tilde{\Omega}_{12}$, and is given by $\tilde{g}_\text{o}=\sqrt{N}g_{\text{o},k}$. 

Next, we consider $\tilde{\Omega}_{23}=\tilde{g}_\mu$:
\begin{multline}
    \tilde{g}_\mu=\bigg(\frac{1}{\sqrt{N}}\sum_i\langle\tilde{1}|\sigma_{13,i}^*\bigg)\bigg(\sum_jg_{\mu,j}\sigma_{23,j}b_j|\tilde{2}\rangle\bigg)
    \\
    =\bigg(\frac{1}{\sqrt{N}}\sum_i\langle\tilde{1}|\sigma_{13,i}^*\bigg)\bigg(\frac{1}{\sqrt{N}}\sum_jg_{\mu,j}\sigma_{13,j}b_j|\tilde{1}\rangle\bigg)
    \\
    =\frac{1}{N}\sum_{i=j}g_{\mu,i}=g_{\mu,k}.
\end{multline}
Similarly, $\tilde{\Omega}_{34}=\Omega_{34,k}$. Hence, the transitions to and from $|1\rangle$ are collectively enhanced, but the others are not. This analysis shows that in our four-level scheme, the optical quantum field is collectively enhanced and the microwave quantum field is not. We note that this is different than the three-level schemes employed in solid-state MOC platforms~\cite{Williamson2014}.

\section{Dipole matrix elements}
We now evaluate the dipole matrix elements for the transitions shown in Fig.~\ref{fig:Figure1}. We begin by calculating the dipole matrix elements and branching ratio from $^3$D$_1$ to $^3$P$_J$ for $^{174}$Yb. Then we numerically calculate the dipole matrix elements of the transitions involving Rydberg atoms. We refer to reference~\cite{Budkerbook} for a detailed description of the angular parts of the DME and the corresponding use of the Wigner-Eckart theorem.

\subsection{$^3$D$_1\rightarrow^3$P$_J$ branching ratios for $^{174}$Yb}
We consider all the decay pathways from the $6s5d$ $^3$D$_1$ $m_J=0$ state to the $6s5p$ $^3$P$_J$ manifold. Note that for $^{174}$Yb there is no nuclear structure since $I=0$. There are six total decay paths: $^3$P$_2$ $m_J=0\,\pm1$; $^3$P$_1$ $m_J=0\,\pm1$; and $^3$P$_0$ $m_J=0$. The latter is the desired decay path, and we consider its relative weight with and without Purcell enhancement from the optical resonator. In the absence of Purcell enhancement, we find a branching ratio to the $^3$P$_0$, $^3$P$_1$, and $^3$P$_2$ manifolds of 20:1.5:1, respectively, such that the probability of emission to $^3$P$_0$ is 89\%. Our system operates in the strong-coupling regime where $\tilde{g}_\text{o}\gg\kappa_\text{e}^\text{o}$ as required for impedance matching, so unfortunately there is no Purcell-enhancement to improve this branching ratio.

\subsection{Approximate numerical calculations of Rydberg transitions}
Matrix elements of alkaline earth atoms can be accurately calculated using only energy levels as experimental input, but require sophisticated many body perturbation techniques~\cite{Porsev2001,Savukov2002}. Good accuracy can also be achieved using numerical analysis based on model potentials with adjustable parameters that are chosen to match experimental energies~\cite{Santra2004}. We show here that greatly simplified calculations based on Coulomb wavefunctions can provide matrix element estimates with moderate accuracy that are helpful for design of experiments. 

The calculation of one-electron jumps is divided into two parts: determination of angular factors and numerical integration 
for reduced matrix elements. We do not expect the calculation to have high accuracy since electron correlation effects are not accounted for. As we will show the matrix element of the low-lying $^3\!P_0 \rightarrow ^3\!D_1$ telecom band transition is calculated with about 40\% error. However, we expect matrix elements involving one or two Rydberg states to have better accuracy since electron correlation effects are relatively small in Rydberg states. 

We assume a quantization magnetic field along the $y$-axis, and we assume the microwave field and the three optical fields are polarized along this axis. In this case, the dipole matrix elements are 
\begin{equation}
d_{21}=\frac{1}{3}\langle ns||er||6p\rangle
\end{equation}
\begin{equation}
d_{32}=-\frac{1}{3}\langle n'p||er||ns\rangle
\end{equation}
\begin{equation}
d_{43}=\frac{1}{3}\langle 5d||er||n'p\rangle
\end{equation}
\begin{equation}
d_{14}=-\frac{1}{3}\langle 6p||er||5d\rangle,
\end{equation}
where the reduced matrix element can be expressed in terms of radial overlap integrals as 
\begin{multline}
\langle n'l_2||r||n l_1\rangle=(-1)^{l_2+(1+l_1+l_2)/2}\sqrt{\text{max}(l_1,l_2)} \\
\int_0^\infty dr\,r^3  R_{n'l_2}R_{nl_1}.
\end{multline}
 
$R_{nl_1},R_{n'l_2}$ are radial wavefunctions. The integral can be evaluated with Coulomb wavefunctions~\cite{Seaton1958} $R_{\gamma,l}$ with $\gamma=n-\mu$ an effective non-integer principal quantum number that depends on the quantum defect $\mu$. To find $\mu$ we use experimental energies of the initial and final states. 

\subsection{Rydberg transitions}
For $\ket{2}$ we need the quantum defect of the $^3\!S$ series. A  recent accurate measurement of the $^1\!S_0$ states gave $\mu=4.278$~\cite{Lehec2018}. The relatively weak electron correlations between ground and  highly excited Rydberg orbitals implies that this value is approximately correct for the triplet state which should have a slightly lower energy and therefore larger quantum defect. The highest $^3\!S_1$ state that has been measured is $6s28s\, ^3\!S_1$~\cite{Camus1980} which has a quantum defect of $\mu_2=4.405$. We will use this value for matrix element estimates.

The energy levels of the $\ket{^3\!P_J}$ series have been measured in~\cite{Aymar1984} together with a multichannel quantum defect analysis of series perturbers. In line with our simplified treatment of matrix elements we extract an effective quantum defect as the value which gives the best root mean square agreement with measured energy values. For $|^3\!P_0\rangle$ we find $\mu_{3,J=0}=4.140$ (matching to energies for $20\le n \le 33$). Using these values for the quantum defects, Table~\ref{tab.nsnpnuJ0} lists the transition frequencies and one-electron reduced transition dipole matrix elements for Rydberg transitions suitable for interfacing with superconducting qubits. Specifically, we tabulate the values for $n=n'=50,\,60,\,70$, and $80$, while we have chosen to focus on $n=n'=60$.

\begin{table}[!t]
\centering
\begin{tabular}{|c|c|c|c|c|}
\hline
$n$ & $\omega_\mu/2\pi~\rm (GHz)$& $\bra{ns}|r|\ket{6p}$& $\bra{np}|r|\ket{ns}$ & $\bra{5d}|r|\ket{np}$     \\
\hline
50 & 18.2& -0.0080& -2900.&-0.0014\\
60 & 10.1& -0.0059& -4310.&-0.0011\\
70 & 6.14& -0.0046&-5990. &-0.00084\\
80 & 4.02& -0.0037 &-7830. &-0.00068\\

\hline
\end{tabular}
\caption{
\label{tab.nsnpnuJ0}Transition frequencies and reduced matrix elements for  Rydberg terms $6snp\, ^3\!S_1$  and $6snp\, ^3\!P_0$. The reduced dipole matrix elements are expressed in units of $e-a_0$. See text for quantum defect values assigned to the participating orbitals.}
\end{table}

\subsection{Fine structure of $^3\!P_J$ levels}
The protocol for microwave to optical conversion uses fields that are detuned from the participating atomic levels. These fields can couple to additional fine structure levels that are not part of the structure shown in Fig. \ref{fig:Figure1}. In order for these additional couplings to be negligible it is necessary that the detunings are small compared to fine structure splittings. The fine structure of the $6s 6p\, ^3\!P_J$ and $6s5d\, ^3\!D_J$ terms is large enough for such off-resonant couplings to be completely negligible. 

However, the Rydberg $6s np\, ^3\!P_J$ splittings are much smaller. If the $J$-dependent quantum defects of the $^3\!P_J$ terms differ by $\epsilon$ this implies an energy difference at principal quantum number $n$ of 
$$
\delta U \simeq |U(n)-U(n-\epsilon)|\simeq \frac{2 U_{Ry} \epsilon}{n^3}.
$$
For ytterbium the difference in quantum defects between $J=0,2$ is 0.225. Assuming the Land\'e interval rule the smallest splitting is between $J=0$ and $1$ which will have $\epsilon=0.225/3 = 0.075$. 
For the highest levels  we are considering $(n=80)$ this implies a splitting of about 680 MHz which is much larger than the $<$MHz-scale Rabi frequencies we are considering. Thus off-resonant coupling to other $J$ levels will be negligible. 

\section{Lifetimes}
There are two mechanisms which limit the lifetime of the atomic system during continuous-wave operation: the finite lifetimes of the intermediate states, and the collisional decay of the atomic sample from the metastable $6s6p$ $^3$P$_0$ state back to the ground state.

\subsection{Spontaneous emission}
The lifetimes of the low lying triplet states of Yb have been accurately measured. The $^3\!P_0$ upper level of the clock transition in bosonic Yb is $> 20~\rm s$. The $^3D_1$ state that participates in the telecom transition has lifetime $\tau=329.3~\rm ns$~\cite{Beloy2012}. The Rydberg lifetimes of the $^3\!S$ and $^3\!P$ series are not well known for large $n$.  In the following we extrapolate low $n$ data to provide rough estimates for high $n$. For the $^3\!S$ series the highest $n$ lifetime measurement that has been reported is $\tau_2=34.3~\rm ns$ at $n=8$\cite{Baumann1985}. With $\left(n^*\right)^3=(n-\mu)^3$ scaling the estimated lifetimes are $\tau_2 = 70, 130, 210, 320, 460~\mu\rm s$ at $n=50,60,70,80,90$.

The $^3\!P$ series are subject to perturbations which cause short lifetimes at particular values of $n$\cite{Zhankui1991}. We can make an approximate estimate for large $n$ values that are not perturbed by extrapolating from a low $n$ non-perturbed lifetime with $\tau_3\sim \left(n^*\right)^3$ scaling. At $n=15$ the $^3P$ lifetime is about $1.2~\mu\rm s$ and within about 10\% in value to the unperturbed $^1P$ lifetime at the same $n$\cite{Zhankui1991}. With $n^3$ scaling this gives lifetimes of $\tau_3 = 90,160,260,400,580~\mu\rm s$ at $n=50,60,70,80,90$.

Finally, due to the finite CW population of the intermediate states $|2\rangle$, $|3\rangle$ and $|4\rangle$ which have lifetimes of $\tau_2$, $\tau_3$, and $\tau_4$, respectively, we must also consider radiative decay. We consider the populations during conversion in both directions. For conversion in the clockwise direction (microwave-to-optical), the steady-state populations for $|2\rangle$, $|3\rangle$, and $|4\rangle$ are given by $P_2^\text{cw}=\tilde{\Omega}_{12}^2/\delta_2^2$, $P_3^\text{cw}=P_2^\text{cw}g_\mu^2/\delta_3^2$, and $P_4^\text{cw}=P_3^\text{cw}\Omega_{34}^2/\delta_4^2$, respectively. For the values given above we have $P_2^\text{cw}\approx0.1$, $P_3^\text{cw}\approx0.1$, and $P_4^\text{cw}\approx0.01$. For conversion in the counter-clockwise direction (optical-to-microwave), the steady-state populations are given by $P_4^\text{ccw}=\tilde{g}_\text{o}^2/\delta_4^2\approx0.1$, $P_3^\text{ccw}=P_4^\text{ccw}\Omega_{34}^2/\delta_3^2\approx0.1$, and $P_2^\text{ccw}=P_3^\text{ccw}g_\mu^2/\delta_2^2\approx0.01$. Note that the populations are the same for the two directions. 

The steady-state lifetimes are $\tau_i=P_i\tau_i$. In the clockwise case, we have $\tau_2^\text{CW}=1.1$ ms, $\tau_3^\text{CW}=1.7$ ms, and $\tau_4^\text{CW}=31$ $\mu$s. In the counter-clockwise case, we have $\tau_4^\text{CCW}=3$ $\mu$s, $\tau_3^\text{CCW}=1.4$ ms, and $\tau_2^\text{CCW}=12$ ms. The conversion time is given by $1/\tilde{\Gamma}_{\text{o}\mu}\approx1.1$ $\mu$s. The decay during conversion in both directions is dominated by $|4\rangle$, which will limit the efficiency to $\approx0.94$ in the clockwise direction and $\approx0.7$ in the counter-clockwise direction. However, these limits are higher than the $\eta\approx0.5$ due to the combination of impedance matching and intrinsic cavity loss.
 
\subsection{Collisional quenching}
Further, we consider the finite lifetime of the ensemble prepared in the $^3$P$_0$ state, as has been measured recently for $^{174}$Yb. The loss per density per time was measured to be $1.3(0.7)\times10^{-11}$ cm$^3$s$^{-1}$~\cite{Franchi2017}. This can be interpreted to mean that a sample of density $\approx8\times10^{10}$ cm$^{-3}$ will have a lifetime of one second. Comparing this to our density of $\approx2\times10^{10}$ cm$^{-3}$ suggests lifetimes of $>4$ s, which is longer than other limiting timescales such as off-resonant scattering of the pump fields or the trapping beam. 

\section{Inhomogeneous broadening}
We must consider various inhomogeneous broadening mechanisms in the ensemble. Since we are interested in a single Rydberg excitation which is collectively enhanced by the $N=150$ atoms, we can neglect Rydberg-Rydberg interactions such as van der Waals~\cite{Weber2015,Zeiher2015,ParisMandoki2017} and dipole-dipole interactions~\cite{Kiffner2016}. Further, our ensemble is relatively dilute, with an average inter-atomic spacing of $d\approx3.6$ $\mu$m. It is expected that the $C_6$ coefficients may be relatively small in the $^3$S$_1$ Rydberg series of Yb, which further reduces Rydberg-Rydberg interactions~\cite{Vaillant2012}. We expect the corresponding interaction shift between atoms is $V(d)=C_6/d^6<\Omega_{12}$. 

Further, we consider shifts due to variation in the position and velocity of atoms in the ensemble. The spatial profile can give rise to inhomogeneous Stark shifts due to the finite size of the pump beams. We estimate the maximum Stark shift variation of $|2\rangle$ over the ensemble to be $|\Delta_2|=\Omega_{12}^2/\delta_2=2\pi\times165$ kHz for the conditions considered above. The inhomogeneous Stark shift from the $\tilde{g}_\text{o}$ field will be similar. For the microwave field $g_\mu$, we note that the volume of the ensemble $V_e$ is in a homogeneous region of the microwave field mode volume, so minimal inhomogeneous broadening will result from this Stark shift. Finally, the Dipole trap can also cause inhomogenous Stark shifts. The maximum Stark shift of a trap of depth 20 $\mu$K is $2\pi\times420$ kHz. The maximum inhomogeneous broadening occurs in the radial direction where the intensity variation in the trap over ($1/e$ radius) $\sigma_R=5$ $\mu$m is significant, for which the inhomogeneous Stark shift of the trap is $|\Delta_\text{trap}|\approx2\pi\times140$ kHz.

We also consider Doppler shifts due to the finite velocity of the atoms in the sample. However, even for a temperature of 10 $\mu$K the Doppler shift is only $\omega_D=2\pi\times20$ kHz, owing partially to the large mass of Yb.  These shifts are all small compared to $\delta_i$ as well as the drive fields that couple all pairs of states. Thus, we neglect them from the analysis above.

\section{Thermal management of optical power}
We analyze the incident power on the side of the microwave resonator plate, and on the aperture of the resonator box. The separation between the resonator plates is $100$ $\mu$m, and the edge of the plate is at $150$ $\mu$m from the waist position of the beam. As discussed above, the pump beams (wavelengths 302 and 386 nm) have waists ($1/e^2$ radii) of $25$ $\mu$m, and the dipole trapping beams (wavelength 760 nm) also have waists of $25$ $\mu$m. The Rayleigh lengths for the pump and trapping beams are $\approx7$ and $\approx3$ mm, respectively. The powers incident on the disk-shaped capacitor plates are below $10^{-20}$ W for both beams, which is negligible compared to the cooling power in $^3$He cryogenic environments.

We further consider the aperture sizes required on the resonator box. The faces of the box along the z-axis are 8 mm from the center. We find that a window of radius 2.5 mm is required for the incident power on the surface to be $<1$ nW for the trapping beams. The incident power from the pump beams is negligible for the pump beams since the Rayleigh length is longer in the UV. However, we must also consider the separation between the two dipole trapping beams at this position, which have a relative angle of $\approx1$ degree. We find their separation to be $\approx2.4$ mm, for which we consider an elliptical aperture as shown in Fig.~\ref{fig:Figure2}a. Such a design has minimal effects on the properties of the microwave resonator.

\bibliographystyle{h-physrev}

\end{document}